\documentclass[runningheads]{llncs}
\usepackage{amsmath}

\usepackage[T1]{fontenc}
\usepackage{graphicx,verbatim}
\usepackage{booktabs}
\usepackage{amsmath}
\usepackage{amsfonts}
\usepackage{comment}
\usepackage{subcaption} 
\usepackage[numbers]{natbib}
\usepackage{placeins}

\newcommand{\class}[1]{\text{class}\left({#1}\right)}
\DeclareMathOperator*{\argmax}{arg\,max}

\begin{document}
\title{This EEG Looks Like These EEGs:\\ Interpretable Interictal Epileptiform Discharge Detection With ProtoEEG-kNN}

\titlerunning{This EEG Looks Like These EEGs}

\author{Dennis Tang\inst{1} \and
Jon Donnelly\inst{1} \and
Alina Jade Barnett\inst{1} \and
Lesia Semenova\inst{2} \and
Jin Jing\inst{3} \and
Peter Hadar\inst{4} \and
Ioannis Karakis\inst{5,6} \and
Olga Selioutski\inst{7} \and
Kehan Zhao\inst{3} \and \\
M. Brandon Westover\inst{3} \and
Cynthia Rudin\inst{1}}
\index{Tang, Dennis}
\index{Donnelly, Jon}
\index{Barnett, Alina Jade}
\index{Semenova, Lesia}
\index{Jing, Jin}
\index{Hadar, Peter}
\index{Karakis, Ioannis}
\index{Selioutski, Olga}
\index{Zhao, Kehan}
\index{Westover, Brandon}
\index{Rudin, Cynthia}
\authorrunning{D. Tang et al.}

\institute{Department of Computer Science, Duke University, USA \and
Microsoft Research, USA \and
Beth Israel Deaconess Medical Center, Harvard Medical School, USA \and
Massachusetts General Hospital, Harvard Medical School, USA \and
Department of Neurology, Emory University School of Medicine, USA \and
Department of Neurology, University of Crete School of Medicine, Greece \and 
Department of Neurology, Stony Brook University, USA \\
\email{Dennis.tang@duke.edu}}
\maketitle 
\begin{abstract}
The presence of interictal epileptiform discharges (IEDs) in electroencephalogram (EEG) recordings is a critical biomarker of epilepsy. 
Even trained neurologists find detecting IEDs difficult, 
leading many practitioners to turn to machine learning for help. While existing machine learning algorithms can achieve strong accuracy on this task, most models are uninterpretable and cannot justify their conclusions. 
Absent the ability to understand model reasoning, doctors cannot leverage their expertise to identify incorrect model predictions and intervene accordingly. To improve the human-model interaction, we introduce ProtoEEG-kNN, an inherently interpretable model that follows a simple case-based reasoning process. ProtoEEG-kNN reasons by comparing an EEG to similar EEGs from the training set and visually demonstrates its reasoning both in terms of IED morphology (shape) and spatial distribution (location). We show that ProtoEEG-kNN can achieve state-of-the-art accuracy in IED detection while providing explanations that experts prefer over existing approaches. 


\end{abstract}

\section{Introduction}

Epilepsy, a chronic neurological disorder characterized by recurring seizures, affects approximately 50 million people worldwide \cite{who_epilepsy_2024}. Epilepsy significantly impairs quality of life, increases risk for injuries, and reduces life expectancy when inadequately managed.
To diagnose epilepsy, clinicians look for electrophysiological events known as interictal epileptiform discharges (IEDs) in electroencephalogram (EEG) recordings. However, identifying IEDs among benign variations in brain activity is difficult, with disagreement being common even among trained neurologists \cite{Jing2020InterraterRO}. To help diagnose epilepsy, clinicians and researchers have recently turned to deep learning methods to create models that reliably detect IED spikes \cite{Diniz2024AdvancingED}. However, despite achieving accurate IED detection, many of these models are uninterpretable -- providing no insight into how decisions are made. This paradigm is problematic because when a practitioner disagrees with a model, there is no way to check the model's reasoning for validity. 

In contrast, interpretable models -- models designed to explain the reasoning behind their decisions -- allow practitioners to assess model predictions and incorporate machine learning insights into the diagnostic process. 
One such model is the Prototypical Part Network (ProtoPNet) \cite{chen2019this}, a family of interpretable neural networks that achieve accuracy on par with black box models on complex tasks. 
However, existing ProtoPNets are ill-equipped to handle the unique challenges of the EEG domain. Specifically, they are unable to handle uncertain labels,
cannot capture the complex interplay between spatial relationships (IED location) and morphological patterns (IED shape) 
that characterize IEDs \cite{Nascimento2022AQA,Kural2020CriteriaFD}, and struggle to learn semantically meaningful prototypes 
due to the extreme variability among IEDs.

\begin{figure}[h!]
    \centering
    \includegraphics[width=0.92\textwidth]{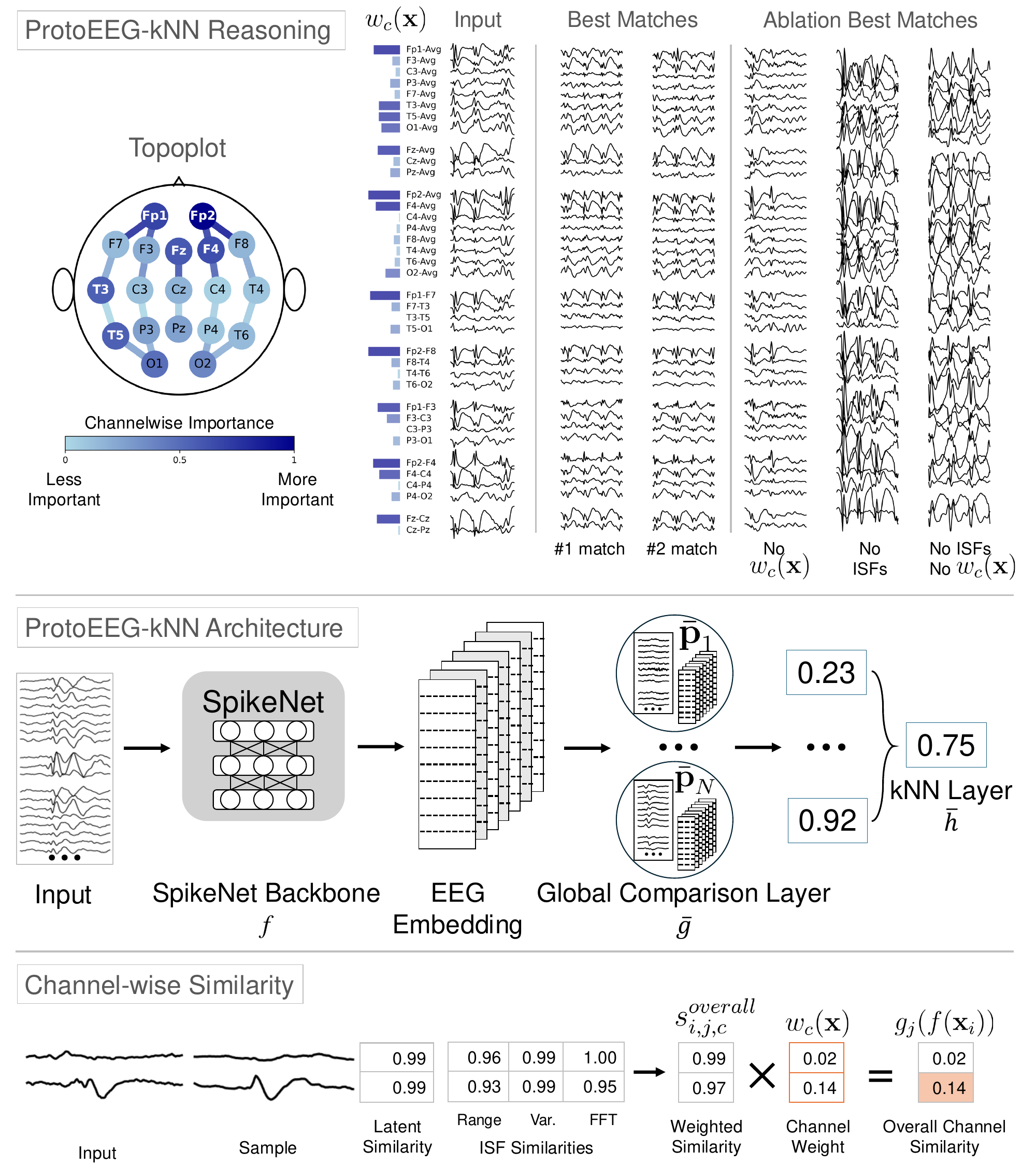}
    \caption{\textbf{Top:} ProtoEEG-kNN reasoning. The topographic map (``topoplot'') highlights important channels as calculated by the channel-wise weights ($w_c(x)$), which are also shown in bars to the left of the input channels. From left to right, we show the input sample, the best two matches selected by our model, and the best matches chosen by each of three ablated models. \textbf{Middle:} ProtoEEG-kNN architecture.  An input is passed through the backbone $f$ to produce a embedding. The Global Comparison Layer $\bar{g}$ computes the similarity between the embedding and each sample in the training set. The final prediction produced by $\bar{h}$ is the average label from the top-k most similar neighbors. \textbf{Bottom:} channel-wise similarity. The similarity along each channel is weighted by $w_c(\mathbf{x})$.} 
    \label{fig:figure1}
\end{figure}

To address these challenges, we introduce ProtoEEG-kNN, an interpretable IED-detection model that achieves state-of-the-art accuracy. 
Our model learns an effective EEG comparison space by training a ProtoPNet with a new similarity metric that incorporates selected interpretable statistical features (ISFs) and specialized spatial reasoning. Once this space is learned, we alter ProtoEEG-kNN to use k-Nearest Neighbors (kNN) reasoning over these learned embeddings, providing intuitive comparisons of the form ``This IED-containing EEG looks like these IED-containing EEGs,'' (Fig. \ref{fig:figure1} (Top)) with coverage over the extreme diversity of IEDs.
Specifically, our contributions are: 
\textbf{(1)} We adapt ProtoPNet into a kNN based probabilistic classification model and update the loss terms to reflect training under uncertain labels.
\textbf{(2)} We define a new similarity metric that aligns our model's notion of EEG similarity with clinical practice by capturing both spike morphology and spatial distribution patterns.
\textbf{(3)} We use channel masking to calculate channel-wise weights that allow the model to prioritize computations on medically relevant channels while revealing the spatial focus of the model's attention across the EEG.

\section{Related Works}


There has been a dramatic increase in interest in IED detection using machine learning models \cite{Diniz2024AdvancingED}, resulting in a wide variety of uninterpretable predictive approaches. Generally, IED detection operates at either the channel-level \cite{Geng2021DeepLF, TjepkemaCloostermans2018DeepLF, Frbass2020AnAI} or by analyzing entire EEGs at once \cite{Kural2022AccurateIO, Jing2020DevelopmentOE, Tveit2023AutomatedIO}.

In computer vision, a large body of work has emerged around interpretable neural networks, based on the Prototypical Part Network (ProtoPNet) \cite{chen2019this}. ProtoPNet provides an interpretable alternative to traditional neural networks by forming predictions using a series of comparisons to learned prototypical parts. A ProtoPNet can explain its predictions by saying ``this image is of class A because it looks like this prototype from class A''. Of particular interest to this work, Ukai et al. \cite{ukai2022looks} introduce ProtoKNN, which performs kNN-style classification over the vector of prototype similarities. This is different from our kNN approach, in which we use a specialized similarity metric to compute the similarity between an input and each training sample.
Several papers have applied ProtoPNet style reasoning to IED detection \cite{Tang2023ProtoEEGNetAI, Gao2023ASD, Gao2024AnIA}. In Gao et al. \cite{Gao2023ASD}, ProtoPNet is applied to IED detection, while Gao et al. \cite{Gao2024AnIA} extends this work to ``multi-scale'' prototypes of varying lengths. However, both are limited to single channel comparisons, thus failing to consider the spatial distribution of spikes, an important factor in how experts identify IEDs \cite{Nascimento2022AQA,Kural2020CriteriaFD}.
In contrast, Tang et al. \cite{Tang2023ProtoEEGNetAI}, represents prototypes as full EEGs, but convolve every channel together, which keeps their model from providing channel-level interpretability. Additionally, Lopez et al. \cite{Lopes2023UsingCS} and Ozcan et al. \cite{Ozcan2019SeizurePI} apply post-hoc methods to explain black-box IED detection models, but these explanations are not necessarily faithful to how a model actually makes decisions, and may be misleading \cite{rudin2019stop, Adebayo2018SanityCF}. 

\section{Methods}\label{sec:methods}
\subsubsection{Notation and Setup.} We denote our training dataset $\mathcal{D}:=\{\mathbf{x}_i, y_i\}_{i=1}^N,$ where $\mathbf{x}_i \in \mathbb{R}^{C \times T}$, $C$ is the number of channels in the EEG and $T$ is the length (1 second sampling 128 Hz), and $y_i \in \{0/v,1/v,\dots,1\}$ (in our case, $v=8$). 
We treat this as a probabilistic classification problem because expert annotators often disagree on labels for this task (in $80.68\%$ of samples in our dataset).
%

Our model architecture is inspired by that of ProtoPNet \cite{chen2019this}, and we train a specialized ProtoPNet to shape the latent space before replacing the learned prototype layer with a kNN module.
During training, the architecture of our model consists of a feature extraction backbone $f: \mathbb{R}^{C \times T} \rightarrow \mathbb{R}^{L \times C}$, prototype layer $g: \mathbb{R}^{L \times C} \rightarrow \mathbb{R}^\textit{M}$, and class-connection layer $h: \mathbb{R}^{\text{M}} \rightarrow [0, 1]$. Here, $L$ and $M$ are the latent dimension and number of prototypes respectively. For our backbone $f$, we use Spikenet, a pre-trained IED classification model. We modify SpikeNet by removing the classifier head and altering the convolution layers to not convolve across EEG-channels, producing embeddings with $C$ separated channels. At the end of training, we replace $g$ and $h$ with kNN style-components $\bar{g}$ and $\bar{h}$, which involves creating a prototype for every training sample and setting $M=N.$ This results in the architecture shown in Fig. \ref{fig:figure1} (Middle).

We now turn to introduce the novel features of our model: A new similarity metric that leverages ISFs, channel-wise weights, and a new kNN layer.



\subsubsection{ISFs and Prototype Similarity.}
In layer $g$, we define each prototype $\mathbf{p}_j \in \mathbb{R}^{L \times C}$ from our set of prototypes $\mathcal{P}_g:=\{\mathbf{p}_j\}_{j=1}^M$ in layer $g$ to represent a complete, 37-channel EEG, and we denote channel $c$ in prototype $j$ with $\mathbf{p}_{j, c} \in \mathbb{R}^{L}$. 
To produce semantically meaningful comparisons, 
we augment the latent cosine similarity with additional comparisons between three ISFs: the range, variance, and fast fourier transform (FFT) of each channel. These comparisons are then aggregated across channels with a weighted sum. We introduce four learnable parameter tensors associated with each prototype $\mathbf{p}_j$: $\mathbf{p}_j^{range} \in \mathbb{R}^C,$  $\mathbf{p}_j^{var} \in \mathbb{R}^C,$ and $\mathbf{p}_j^{\mathit{fft}} \in \mathbb{R}^{C\times T},$ where each entry along the $C$ dimension corresponds to the relevant statistic computed over each channel. 
This yields  four similarity terms: $s^{latent},$ $s^{range}$, $s^{var}$, and $s^{\mathit{fft}}$, where the superscript defines which set of features the similarity scores are computed along.
We define the similarity between a single channel $c$ of input $i$ and prototype $j$ along each measure as:
{\allowdisplaybreaks
\begin{align*}
    &s_{i,j, c}^{latent} = \frac{f_{c}(\mathbf{x}_{i}) \cdot \mathbf{p}_{j, c}  }{\|f_{c}(\mathbf{x}_{i})\|_2\|\mathbf{p}_{j, c}\|_2},     &s_{i,j, c}^{\mathit{fft}} = \frac{c_{\mathit{fft}}}{ \| |\mathbf{p}^{\mathit{fft}}_{j,c}| -  |FT(\mathbf{x}_{i, c})| \|_{2} + \epsilon},\\
    &s_{i,j,c}^{var} = 1 - \left|\frac{Var(\mathbf{x}_{i, c}) - Var(p^{var}_{j,c})}{V_{max} - V_{min} + \epsilon} \right|, 
    &s_{i,j, c}^{range} = 1 - \left|\frac{R(\mathbf{x}_{i,c}) - R(p^{range}_{j,c})}{R_{max} - R_{min} + \epsilon}\right|,
\end{align*}
}
where $f_{c}(\mathbf{x}_{i}) \in \mathbb{R}^L$ denotes the latent representation of the $c$-th channel in $\mathbf{x}_i$, $Var(\cdot)$ is the variance, $R(\cdot)$ is the range, $FT(\cdot)$ is the fourier transform, $\epsilon$ and $c_{\mathit{fft}}$ are constants for numerical stability, and $V_{min},$ $V_{max},$ $R_{min},$ and $R_{max}$ denote the minimum variance, maximum variance, minimum range, and maximum range across all channels in the training set respectively.
An overall similarity score between two channels is calculated as:
    $s_{i,j, c}^{overall} = \lambda_1s_{i,j, c}^{latent} + \lambda_2s_{i,j, c}^{range} + \lambda_3s_{i,j, c}^{var} + \lambda_4s_{i,j, c}^{\mathit{fft}},$
where $\lambda_i := sm(\lambda'_1, \lambda'_2, \lambda'_3, \lambda'_4)$ for learned parameters $\lambda'_1, \lambda'_2, \lambda'_3, \lambda'_4,$ and $sm$ denotes the softmax function.

\subsubsection{Channel-wise Weights.}
To focus the model's similarity comparisons along relevant channels and to provide channel-level interpretability, we calculate a channel-wise weight for every channel in the input. We use a leave-one-channel-in masking approach 
 and define the weight function ${w_{c}} : \mathbb{R}^{C \times T} \to \mathbb{R}$ such that
${w}_c({\mathbf{x}_{i}}) = \frac{\Tilde{w}_{c}({\mathbf{x}_{i}})}{\sum_{c \in C} \Tilde{w}_{c}({\mathbf{x}_{i}})}, 
    \Tilde{w}_{c}({\mathbf{x}_{i}}) = h_{spikenet}(f([\mathbf{0}^{c-1 \times T}; \mathbf{x}_{i,c}; \mathbf{0}^{C-c \times T})),$
where $f$ is the backbone, $h_{spikenet}$ is the classifier head of SpikeNet, $\mathbf{0}^{A \times B}$ denotes an $A \times B$ dimensional matrix of zeroes, and $;$ indicates concatenation. 
Each weight ${w}_c({\mathbf{x}_{i}})$ assigns a relative importance to the similarity score along channel $c$, 
yielding an overall similarity score:
    $g_j(f(\mathbf{x}_i)) = 
    \sum_{c = 1}^{C} w_c(\mathbf{x}_{i})s_{i,j,c}^{overall}$ (Fig. \ref{fig:figure1}(Bottom)).
%
%
Given our similarity function, we focus next on model training.

\subsubsection{Weighted Loss Terms.} 
We train our model to produce well calibrated predictions using the binary-cross entropy loss  $\mathcal{L}_{bce} = -y_i \log(\hat{y}_i) - (1 - y_i)\log(1-\hat{y}_i)$. This way, we can retain the primary function of IED-classifcation with the added benefit of calibrating our model to also match the vote proportions.

Moreover, we adapt the loss terms (Cluster, Separation, Orthogonality) from ProtoPNet to handle uncertain labels. 
Let $cos(\mathbf{p}_j, \mathbf{p}_{j'}):= \frac{vec(\mathbf{p}_j) \cdot vec(\mathbf{p}_{j'})}{\|vec(\mathbf{p}_j)\|_2 \|vec(\mathbf{p}_{j'}) \|_2}$ denote the cosine similarity between two prototypes, where $vec(\mathbf{p}_j)$ denotes the vectorization of $\mathbf{p}_j$. 
We define the loss across a batch as:
{\allowdisplaybreaks
\begin{align*}
&\mathcal{L}_{ortho} = \sqrt{\sum_{j=1}^M \sum_{j'=1}^M \mathbf{1}_{[j\neq j']}cos^2(\mathbf{p}_j, \mathbf{p}_{j'})} + \sqrt{\sum_{j=1}^M \sum_{j'=1}^M \mathbf{1}_{[j\neq j']}cos^2(\mathbf{p}^{\mathit{fft}}_j, \mathbf{p}^{\mathit{fft}}_{j'})}~~,\\
&\mathcal{L}_{clst} = -\frac{1}{B} \sum_{i=1}^B \max_{j^\dagger: \text{class}(j^\dagger)=y_i} y_{i}g_{j^\dagger}(f(\mathbf{x}_i)),\\ 
&\mathcal{L}_{sep} = \frac{1}{B} \sum_{i=1}^B  g_{j^*}(f(\mathbf{x}_i)) \cdot  |\class{j^*} - y_{i}|, \text{where } j^* = \argmax_{j: \text{class}(j)\neq y_i} g_{j}(f(\mathbf{x}_i)),
\end{align*}
}
where $\mathbf{1}_{[\cdot]}$ denotes the indicator function, 
$\class{j}$ is the class associated with prototype $j$, and $B$ is the batch size. Finally, 
we add a regularization loss $\mathcal{L}_{CoefReg} = \lambda_{1} - min(\lambda_{2},\lambda_{3},\lambda_{4})$ to train balanced coefficients for ISFs.

We minimize the overall loss function $\mathcal{L}_{overall}:=\kappa_1\mathcal{L}_{bce} + \kappa_2\mathcal{L}_{ortho} + \kappa_3\mathcal{L}_{clst}+ \kappa_4\mathcal{L}_{sep} + \kappa_5\mathcal{L}_{CoefReg},$ where each $\kappa$ is a scalar hyperparameter, using Adam optimization. 
We denote the model $h \circ g \circ f$ as ``EEG ProtoPNet'' and train according to the regime described in \cite{chen2019this} to produce a well-structured latent space when combined with our ISFs. Training lasted 200 epochs and stopped early if validation accuracy did not improve for two consecutive project epochs. 

\subsubsection{kNN Replacement Step.}
Once the EEG ProtoPNet training has converged, we replace the learned prototype layer $g$ with a Global Comparison Layer $\bar{g}:\mathbb{R}^{L \times C} \to \mathbb{R}^N$ and the linear layer $h$ with a kNN comparison layer $\bar{h}: \mathbb{R}^N \to [0, 1]$. This is our final model, ``ProtoEEG-kNN.''
The Global Comparison Layer $\bar{g}$ can be thought of as a prototype layer in which every training sample is a prototype. Formally, we set $\mathbf{\bar{p}}_i := f(\mathbf{x}_i)$, ${\bar{p}}_{i,c}^{range} := R(\mathbf{x}_{i,c})$,  ${\bar{p}}_{i,c}^{var} := Var(\mathbf{x}_{i,c})$,  and $\mathbf{\bar{p}}_{i,c}^{\mathit{fft}} := FT(\mathbf{x}_{i,c})$ for $i \in \{1, 2, \hdots, N\},$ and $\bar{g}$ operates as a prototype layer with prototypes $\mathcal{P}_{\bar{g}}:=\{\bar{\mathbf{p}}_i\}_{i=1}^N$. This makes $\bar{g}_j(f(\mathbf{x}_i))$ the similarity between the $j$-th training sample and input $\mathbf{x}_i$, using the weighted similarity metric defined previously.
The kNN layer $\bar{h}$ is then formalized as 
    $\bar{h}\circ\bar{g}\circ f(\mathbf{x}_i) := \frac{1}{k}\sum_{j' \in \text{topk}(\bar{g}(f(\mathbf{x}_i)))} y_{j'},$
where $\text{topk}$ returns the $k$ largest indices in a vector and $y_j$ denotes the label of the $j$-th training sample. ProtoEEG-kNN is therefore the composition $\bar{h} \circ \bar{g} \circ f$.
In Section \ref{sec:results}, we demonstrate this kNN replacement step substantially improves both accuracy and interpretability.

\section{Results}\label{sec:results}
We train and evaluate ProtoEEG-kNN using a dataset of 16,499 EEGs labeled by 8 annotators. Participants were recruited from three settings: intensive care unit (n = 446), routine / outpatient EEG (n = 1,161), and epilepsy monitoring unit (n = 104). The data consists of 841 males (mean age = 36.56 years) and 921 females (mean age = 36.92 years). The data was split into 12,411 training, 2,151 validation, and 1,937 test samples, with no patient overlap between sets. This ensures that input samples are compared only with EEGs from other patients. Samples are arranged in standard, 37-channel, ``double-banana'' format \cite{Jadeja2021HowTR}, were filtered (60-Hz notch, 0.5-Hz high-pass), and re-sampled to 128 Hz. 
Following the annotation procedure in Jing et al. \cite{Jing2020DevelopmentOE}, for each EEG sample, 8 subspecialist physicians independently annotated whether they observed an IED. 

ProtoEEG-kNN was trained on a Nvidia P100 GPU for $\sim5$ clock hours. Class-balanced sampling was used during training and $k$ was set to 10 in $\bar{h}$. We now evaluate ProtoEEG-kNN's accuracy, assess its match-quality, and ablate its novel components.

\subsubsection{ProtoEEG-kNN is Accurate.} We evaluate model performance using binary classification accuracy, AUROC, and $R^{2}$. For binary classification and AUROC, we assign a sample to the positive class if $y_{i} \geq 0.5$. On the held-out test set, we evaluate ProtoEEG-kNN, SpikeNet, kNN over the FFT of EEG samples, kNN over the ISFs of EEG samples, Deep kNN \cite{Zhuang2020DeepKF}, and EEG ProtoPNet.

The optimal weighting coefficients for kNN over the ISFs were determined on the validation set by evaluating every combination of coefficients that sum to 1 in increments of 0.1. For Deep kNNs, we train the latent space of SpikeNet and copy Deep kNN's exact hyper-parameter and optimization configuration. As shown in Table \ref{table:exp1} (Top), ProtoEEG-kNN substantially outperforms existing models for this task in terms of binary classification, AUROC and $R^2$.



\begin{table}[t]
\footnotesize
    \centering
    \begin{tabular}{lccc}  
        \toprule
        Method & Binary Accuracy & AUROC & $R^{2}$ \\
        \midrule
        SpikeNet & 77.12 & 0.844 & 0.429 \\
        kNN over FFT & 70.72 & 0.720 & 0.209 \\
        kNN over ISFs  & 74.39 & 0.733 & 0.210 \\
        Deep kNN \cite{Zhuang2020DeepKF} & 77.16 $\pm$ 0.01 & 0.805 $\pm$ 0.007 & 0.341 $\pm$ 0.019 \\
        EEG-ProtoPNet & 80.24 $\pm$ 0.36 & 0.866 $\pm$ 0.006 & 0.207 $\pm$ 0.019  \\
        ProtoEEG-kNN (ours) & \textbf{81.15 $\pm$ 0.29} & \textbf{0.876 $\pm$ 0.000} & \textbf{0.529 $\pm$ 0.007} \\
        \bottomrule 
        \toprule
        \textbf{Ablations} & & & \\
        ~~Remove $w_{c}$ & 80.74 $\pm$ 0.08 & 0.878 $\pm$ 0.002 & 0.536 $\pm$ 0.003 \\ 
        ~~Remove ISFs & 80.91 $\pm$ 0.00 & 0.878 $\pm$ 0.001 & \textbf{0.538 $\pm$ 0.005} \\
        ~~Remove $w_{c}$ \& ISFs & 81.09 $\pm$ 0.61 & \textbf{0.885 $\pm$ 0.004} & 0.531 $\pm$ 0.027 \\
        ProtoEEG-kNN (complete) & \textbf{81.15 $\pm$ 0.29} & 0.876 $\pm$ 0.000 & 0.529 $\pm$ 0.007 \\
        
        \bottomrule \\
    \end{tabular}
    \caption{ Performance of ProtoEEG-kNN compared to baselines (\textbf{Top}) and ablated models (\textbf{Bottom}). For models that required additional training, we train with 3 different random seeds and report mean and standard deviation. 
    }
    \label{table:exp1}
\end{table}

\begin{figure}[t]
    \centering
    \begin{subfigure}{\textwidth}
        \centering
        \footnotesize
        \renewcommand{\arraystretch}{1.2}
        \begin{tabular}{lccc}
            \toprule
            Method & Plackett-Luce Weight & Best-Match Frequency \\
            \midrule
            Random & 0.128 (0.111, 0.144) & 0.104 \\
            EEG-ProtoPNet & 0.078 (0.065, 0.088) & 0.052 \\
            Deep kNN  & 0.333 (0.298, 0.368) & 0.370 \\
            ProtoEEG-kNN & \textbf{0.462 (0.427, 0.501)} & \textbf{0.474} \\
            \bottomrule
        \end{tabular}
    \end{subfigure}

    \begin{subfigure}{\textwidth}
        \centering
        \includegraphics[height=1.5in]{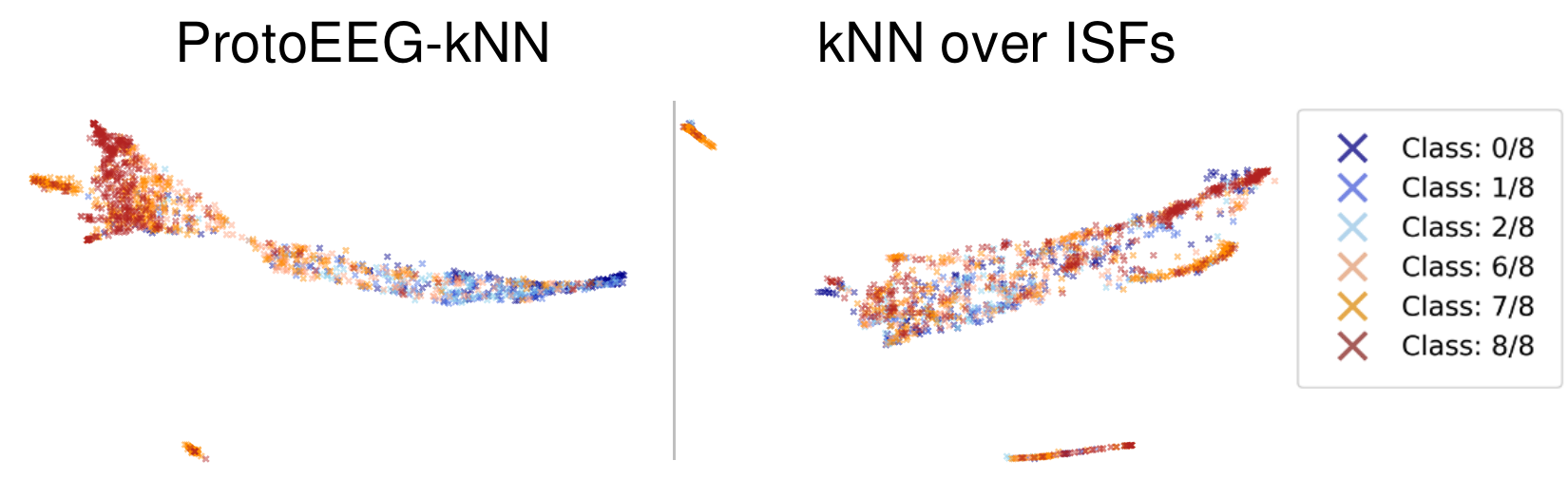}
    \end{subfigure}
    
    \caption{\textbf{Top:} User Study Results. Bootstrapping with 1,000 iterations was used to calculate the mean and 95\% confidence interval for Plackett-Luce weights. \textbf{Bottom:} PaCMAP visualization of the test set comparison spaces of ProtoPNet-kNN (left) and kNN over ISFs (right). Neighborhoods in high-dimensional space are preserved in two-dimensional PaCMAPs.
    }
    \label{fig:figure2}

\end{figure}

\subsubsection{ProtoEEG-kNN produces good matches.} 
To demonstrate that ProtoEEG-kNN produces quality matches that align with medical intuition, we conducted a user study with four board-certified neurologists (with 2-16 years of clinical experience) and one clinical neurophysiology/EEG fellow. Experts were shown 100 `reference' EEG samples from the test set and ranked the similarity of four `candidate' matches. Three candidates were the top matches identified by ProtoEEG-kNN, Deep kNN, and EEG-ProtoPNet, while the fourth was a randomly selected sample sharing the reference's classification label.
For each ranking, the order of candidates was randomized and the selection method was hidden. 
We restricted reference samples to have label  $\geq0.75$ to ensure clear IED patterns for matching.

To quantify each model's match quality, we used best-match frequency and Plackett-Luce model weights. Best-match frequency indicates how often each model was ranked first, while Plackett-Luce weights consider the full ranking distribution and 
represents the probability each model provides the best match \cite{Hunter2003MMAF}. Across both metrics, ProtoEEG-kNN produces matches that align the closest with expert opinion (Fig. \ref{fig:figure2} (Top)). 


We also qualitatively evaluate the comparison space of our model by using the dimension reduction tool PaCMAP \cite{Wang2020UnderstandingHD} to visualize the distribution of the test set under ProtoEEG-kNN’s similarity metric. 
Relative to the comparison space based on the kNN over ISFs' similarity metric, ProtoEEG-kNN learns more distinct and well-separated classes (Fig. \ref{fig:figure2} (Bottom)).

\subsubsection{Ablations.}
Finally, we evaluate ProtoEEG-kNN's performance without channel-wise weights and ISFs (Table \ref{table:exp1} (Bottom)). The inclusion of channel-wise weights and ISFs marginally effects binary classification ($\uparrow$ 0.06\%), AUROC ($\downarrow$ 0.0084), $R^2$ ($\downarrow$ 0.0022), while resulting in much closer matches (Fig. \ref{fig:figure1} (Top)). 

\section{Conclusion}
We introduced ProtoEEG-kNN, an interpretable model for IED detection that achieves state-of-the-art performance while providing interpretable reasoning for its decisions in the form of ``This EEG looks like these EEGs''. In addition to being interpretable, our model's kNN layer, similarity metric, and channel-wise weights scores constrain it to reason in a way that aligns with clinical intuition about spike morphology and spatial distribution, as shown through our user study. 
While ProtoEEG-kNN demonstrated promising results, future work should externally validate ProtoEEG-kNN using different patient populations to confirm its generalizability. Nonetheless, ProtoEEG-kNN offers a promising path forward for the integration of machine learning into clinical practice.
%
%
\bibliographystyle{splncs04}
\bibliography{bibliography}

\appendix

\section{User Study}

For our user study, our experts consisted of 5 physicians (4 MDs, 1 DO). One of our experts was completing a clinical neurophysiology/EEG fellowship while the remaining four were board-certified practitioners with 2, 12, 15, and 16 years of clinical practice. Three of the experts hold professorships at universities in the United States.

An example of a user study question is shown below. 

\begin{figure}[htbp]
    \centering
    \includegraphics[width=0.8\textwidth]{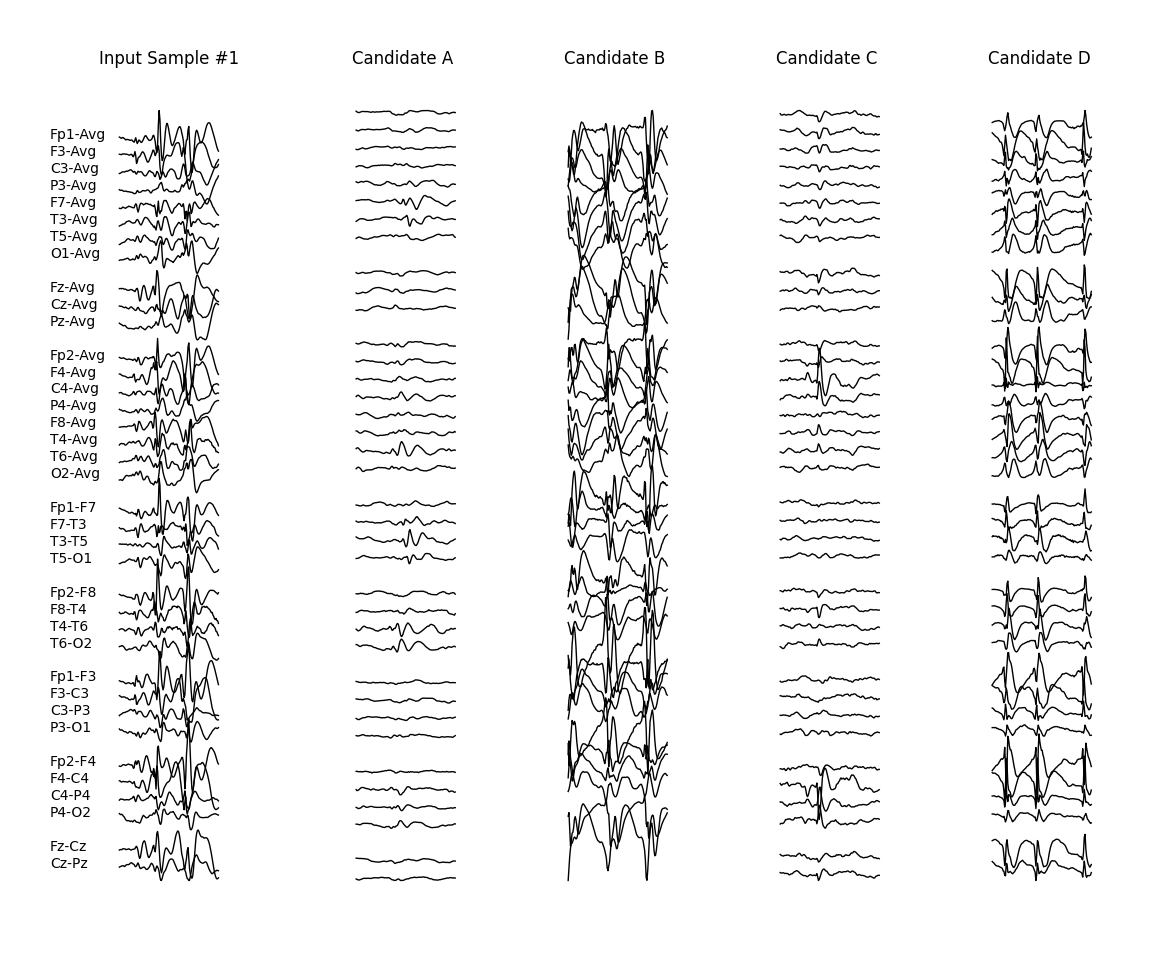}
    \caption{ \textbf{Example of A User Study Ranking}. Experts are asked to rank the similarity of each candidate to the input samples for 100 different inputs. The survey was conducted on a secure Qualtrics platform.}
    \label{fig:user_study}
\end{figure}

\FloatBarrier
\section{Local Analysis}
\FloatBarrier

``Local'' interpretability explains why a decision was made for a single example. This is in contrast to ``global'' interpretability, which explains the overall decision behavior, not linked to any particular example. The following figures show a local analysis for several examples. The classification decision of the left EEG sample is explained by its similarity to the right five samples.

\begin{figure}[htbp]
    \centering
    \includegraphics[width=0.8\textwidth]{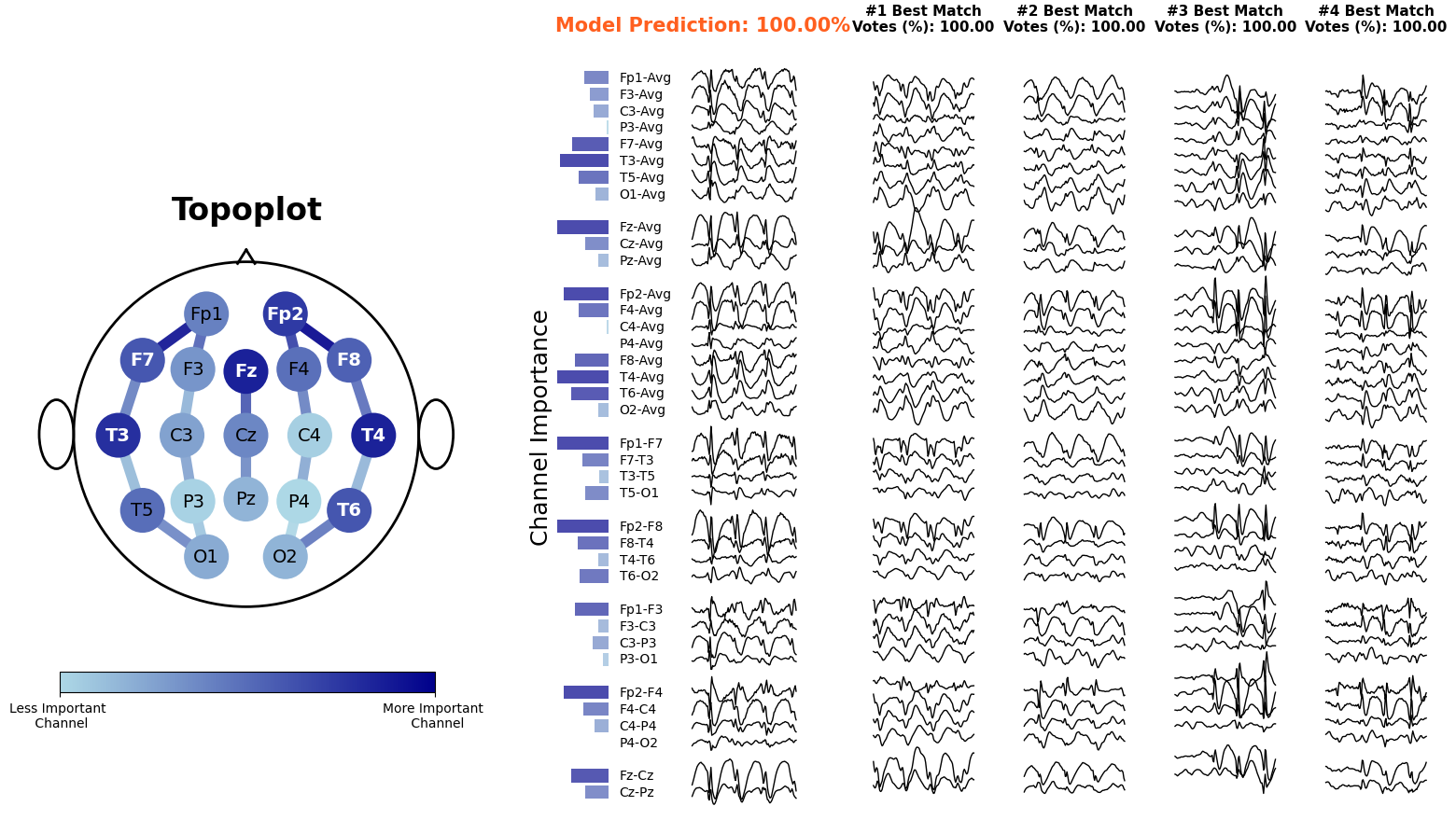}\\
    \includegraphics[width=0.8\textwidth]{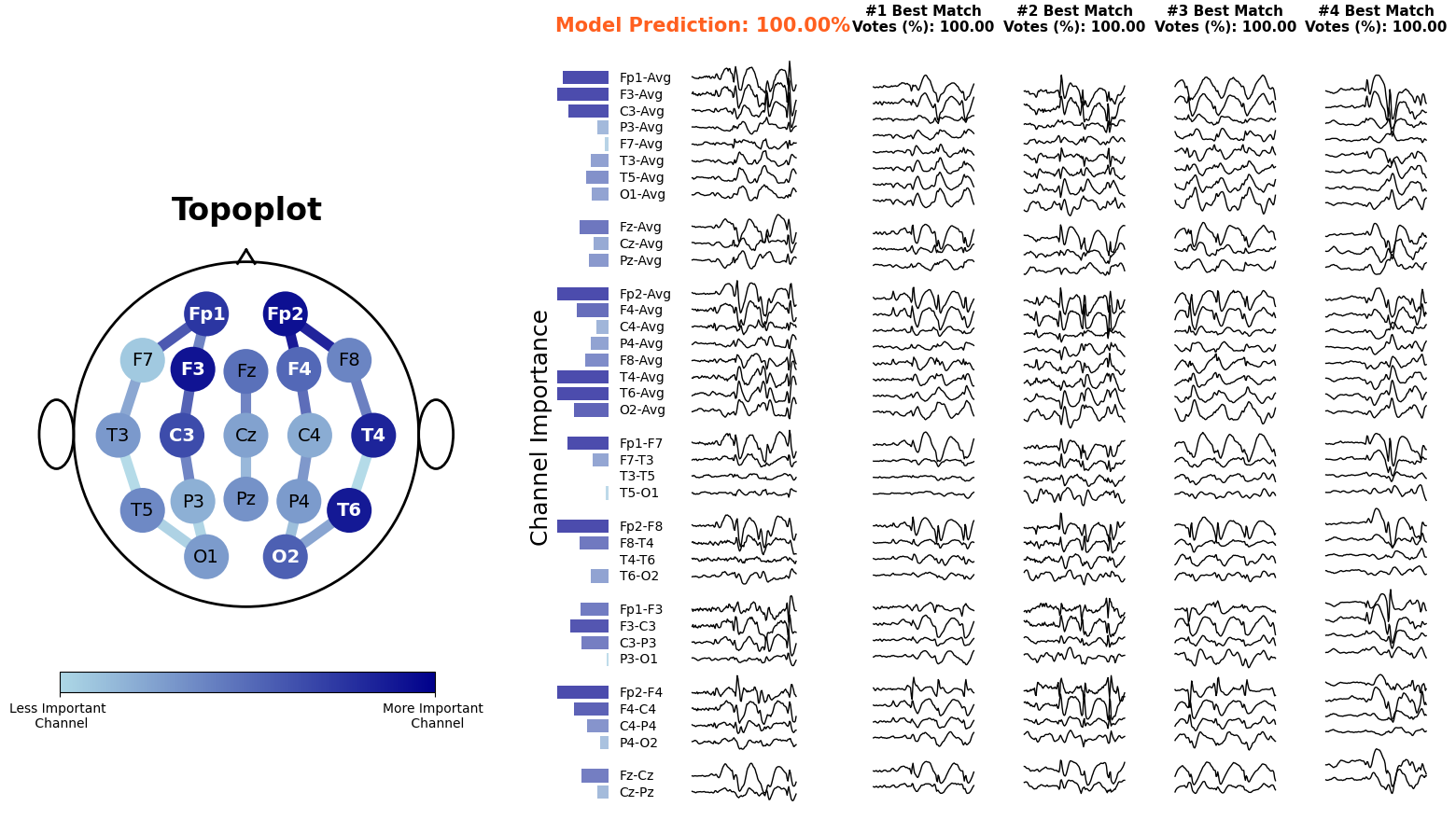}\\
    \includegraphics[width=0.8\textwidth]{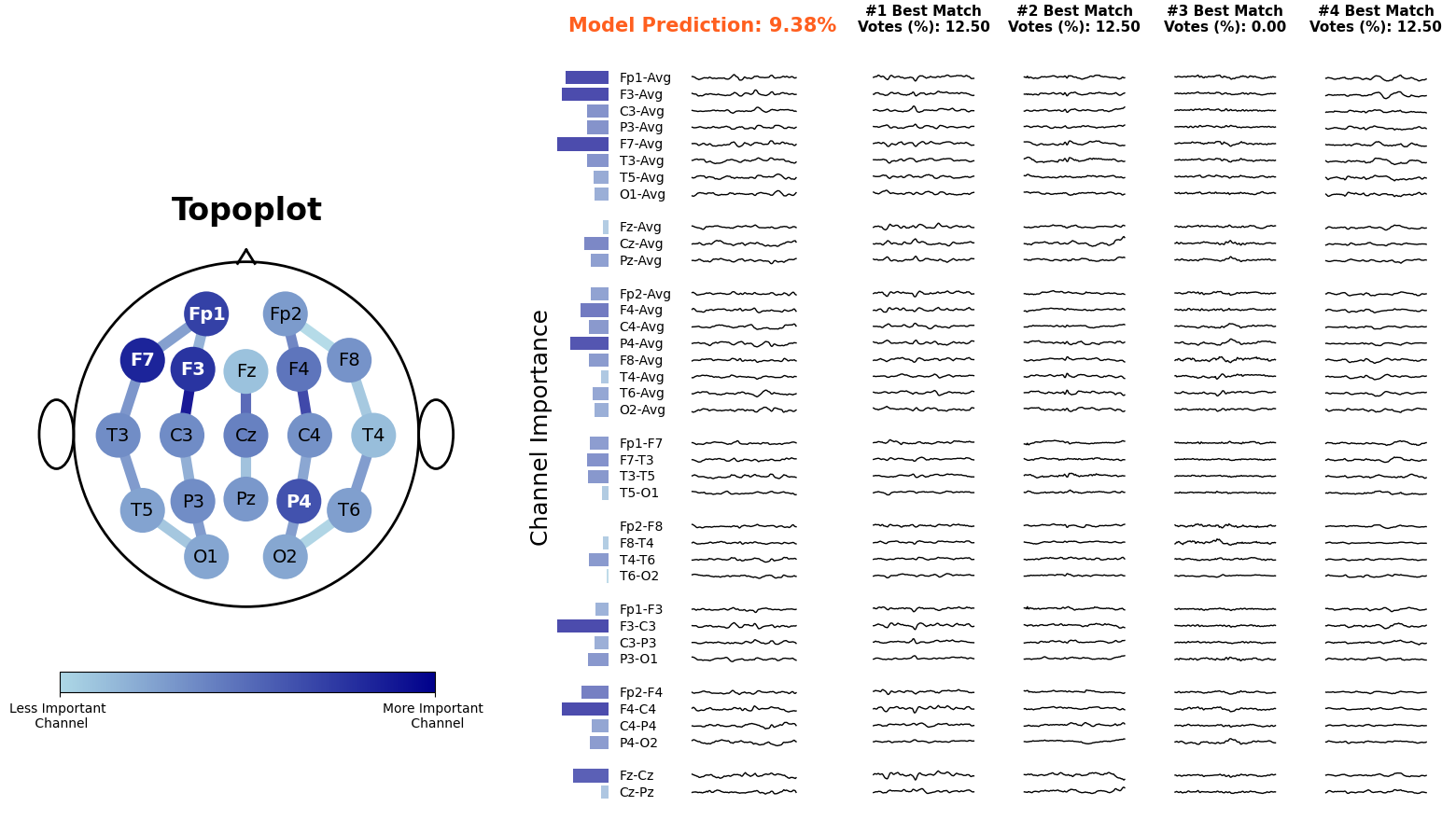}
    \caption{\textbf{Examples of Local Analysis}}
    \label{fig:local_analysis_1}
\end{figure}

\begin{figure}[htbp]
    \centering
    \includegraphics[width=1.0\textwidth]{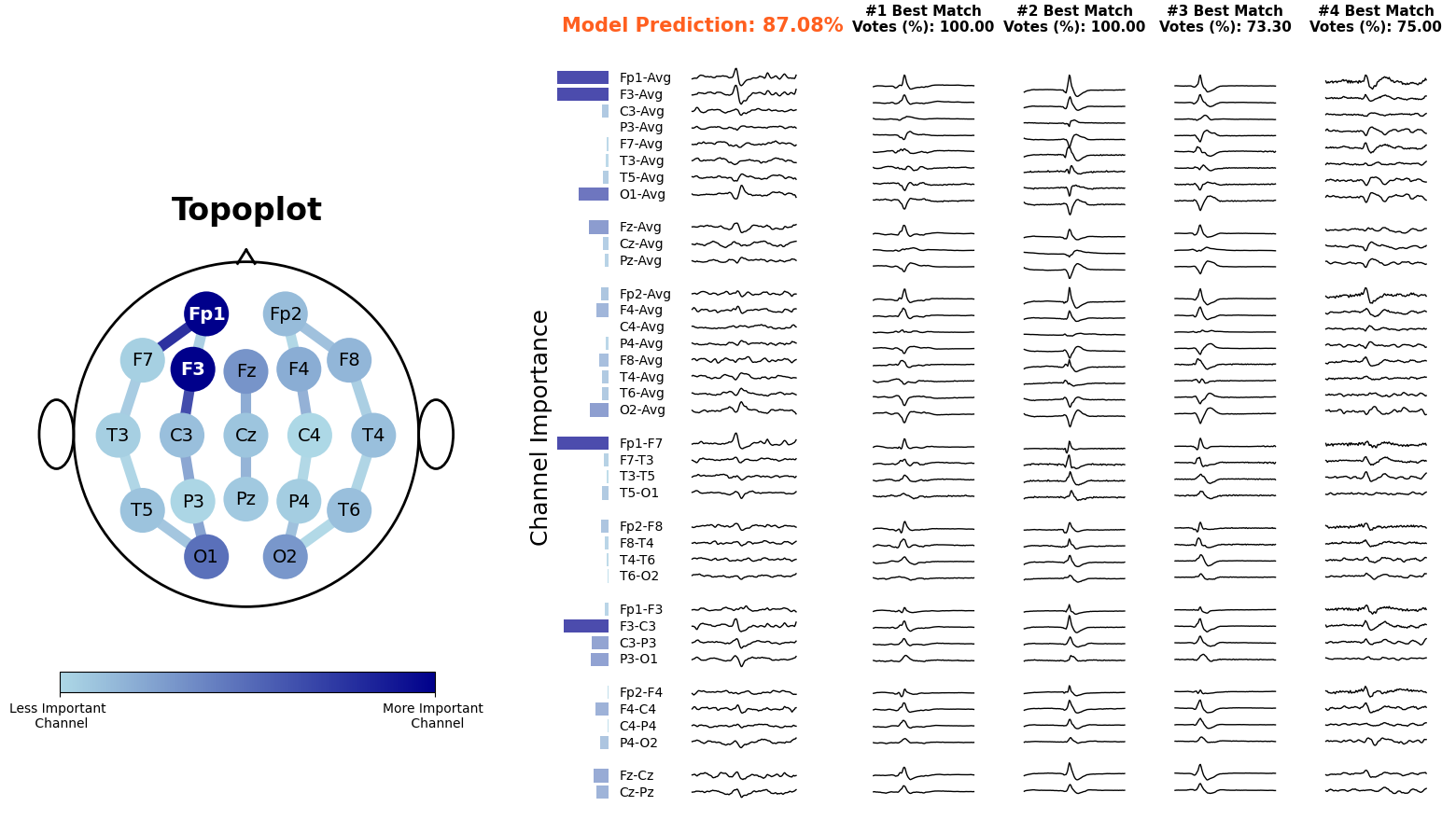}\\
    \includegraphics[width=1.0\textwidth]{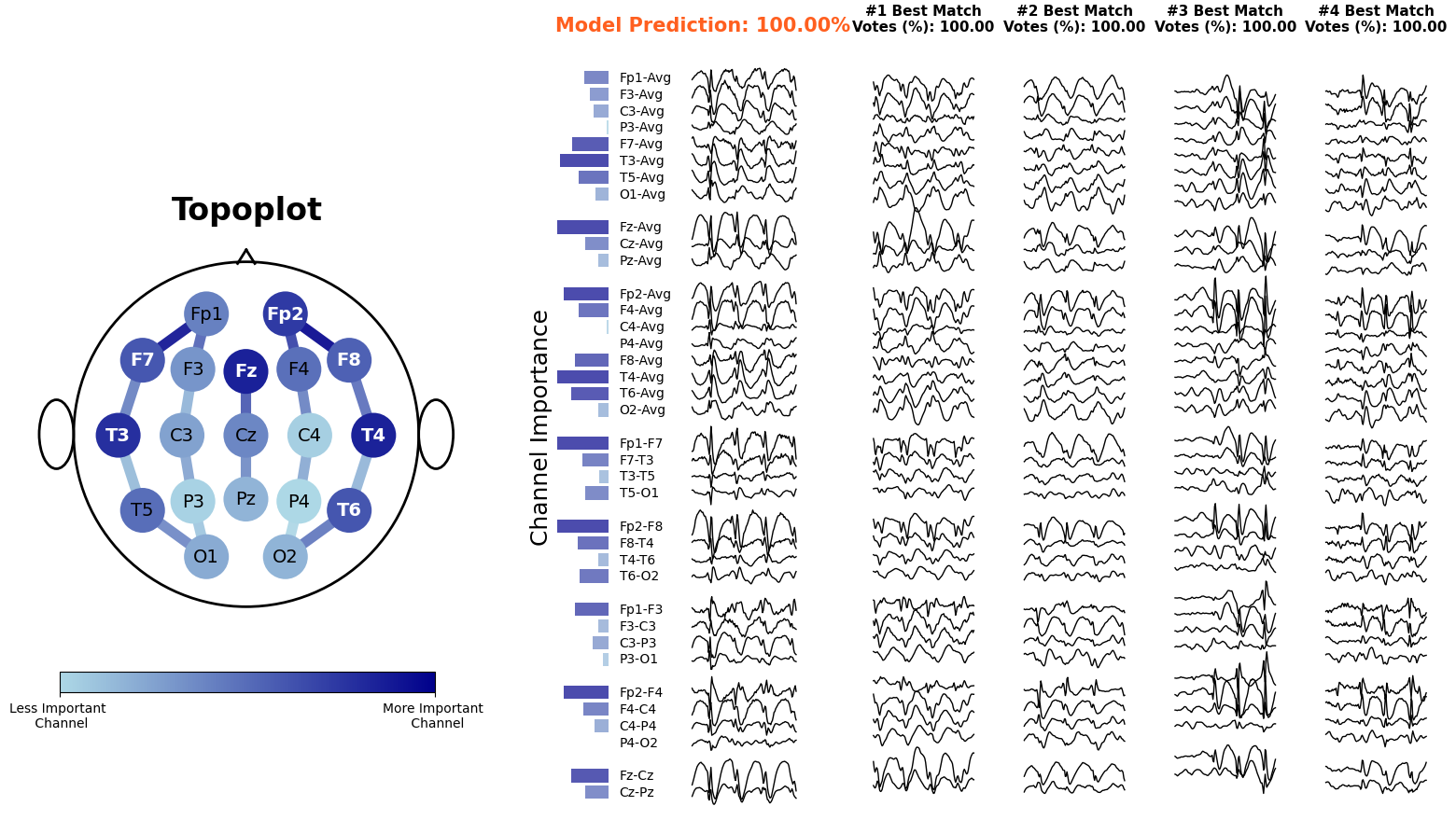}\\
    \includegraphics[width=1.0\textwidth]{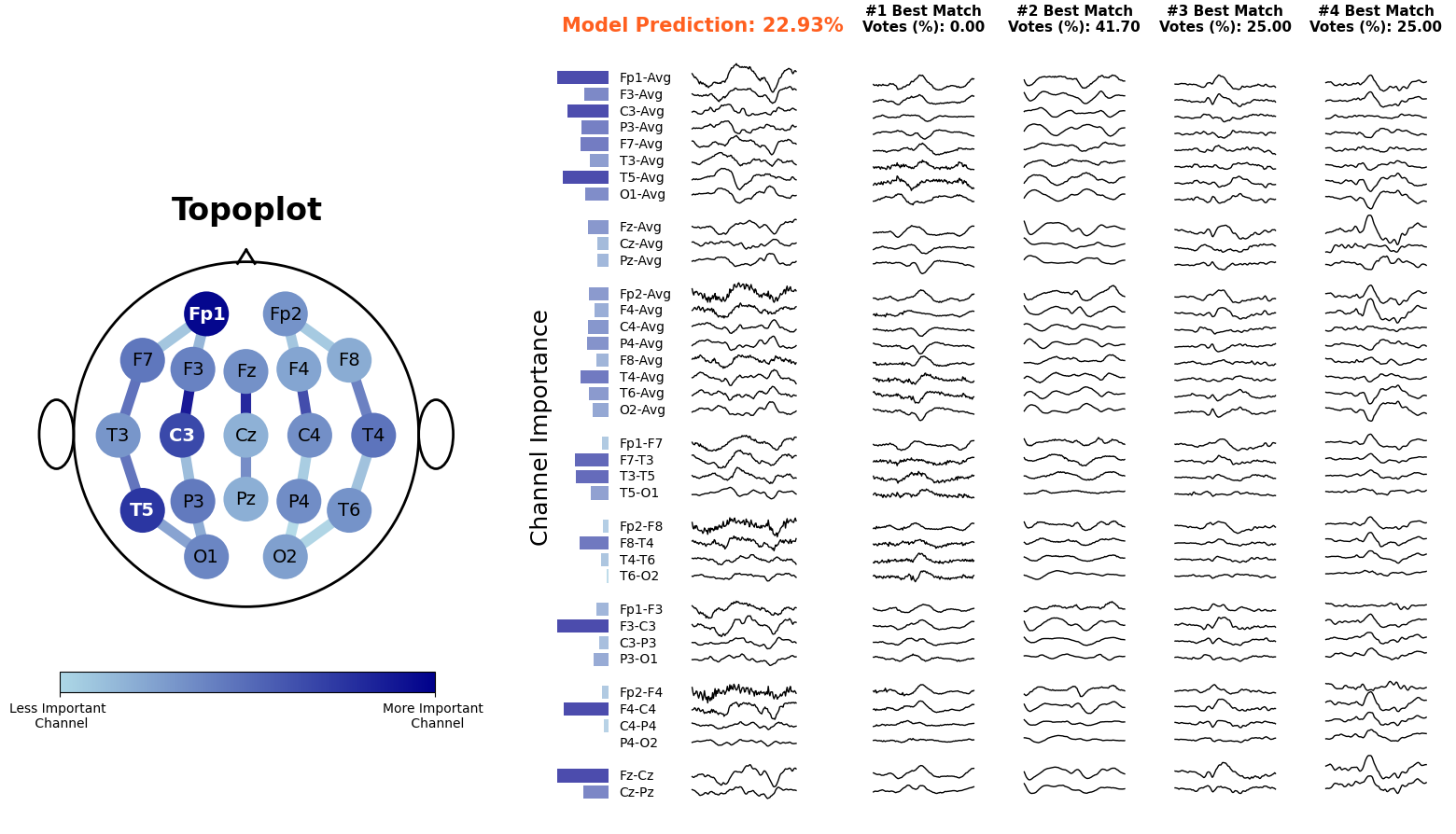}
    \caption{\textbf{More Examples of Local Analysis}}
    \label{fig:local_analysis_2}
\end{figure}

\FloatBarrier
\section{Ablations}
\begin{figure}[htbp]
    \centering
    \includegraphics[width=0.95\textwidth]{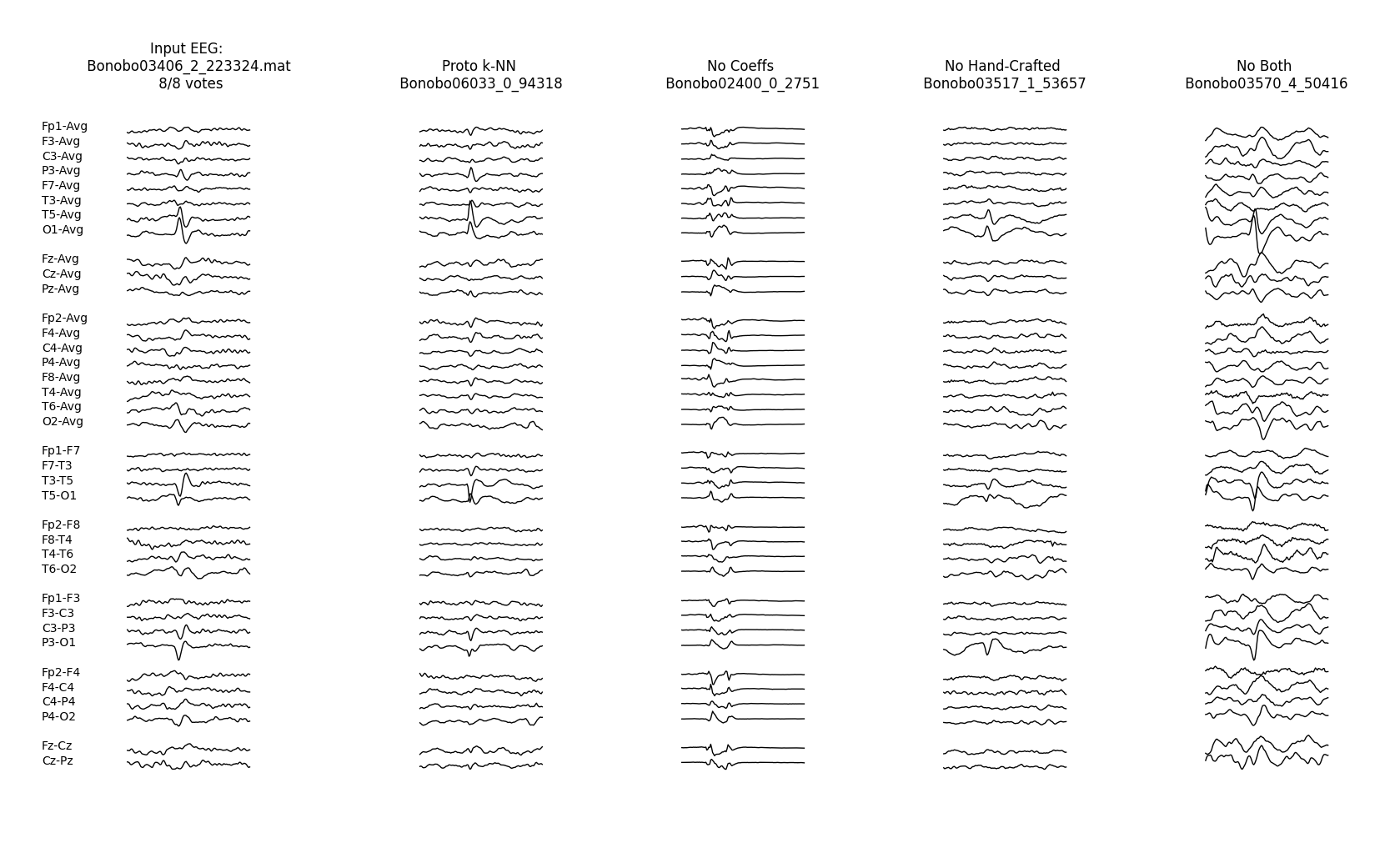}\\[-3ex]
    \includegraphics[width=0.95\textwidth]{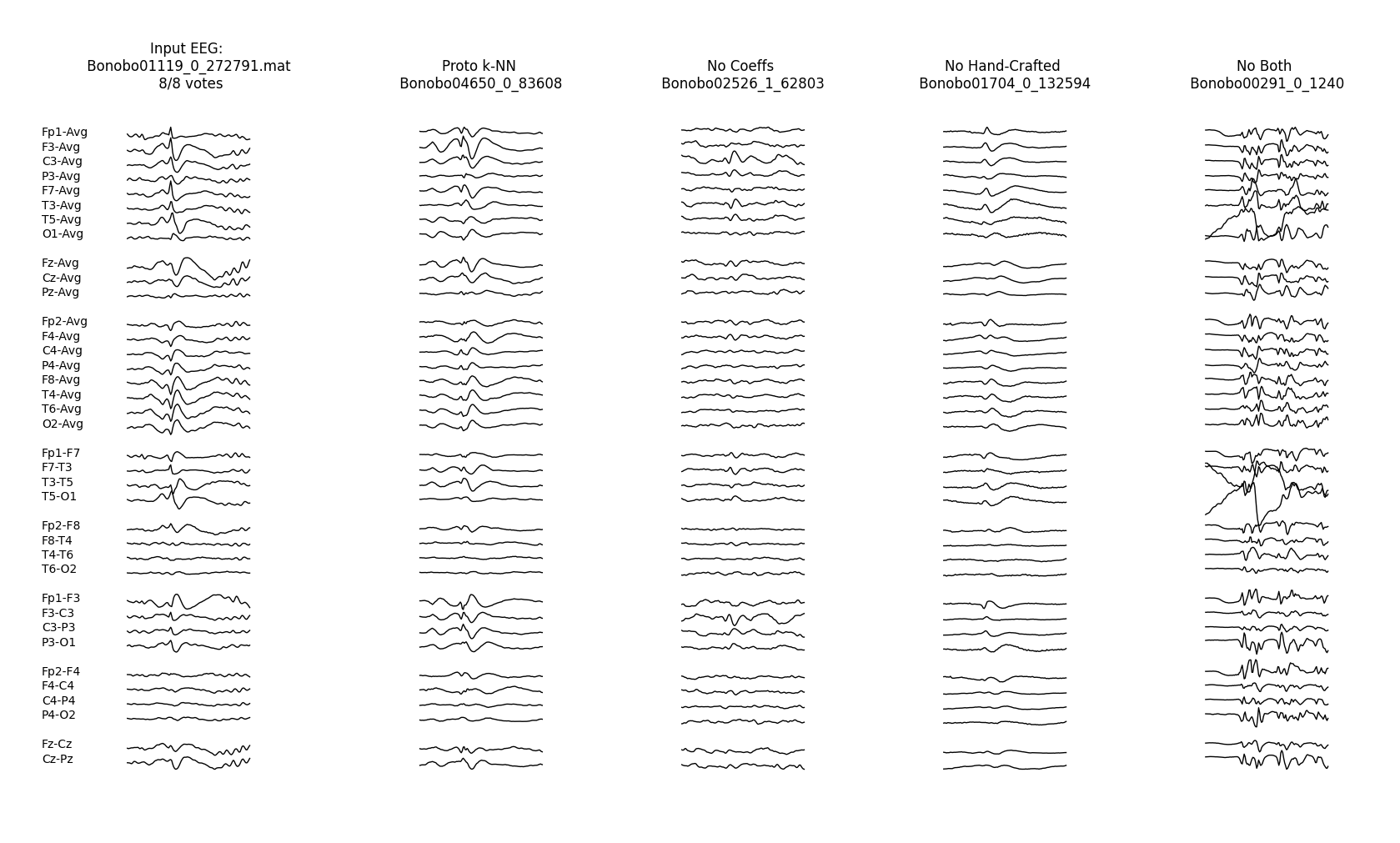}
    \caption{\textbf{Model Ablations Nearest Neighbors.}}
    \label{fig:ablations}
\end{figure}

\end{document}